# A quantum degenerate Bose-Fermi mixture of $^{41}$K and $^6$Li

Yu-Ping Wu[1,2,3], Xing-Can Yao[1,2,3,4], Hao-Ze Chen[1,2,3], Xiang-Pei Liu[1,2,3], Xiao-Qiong Wang[1,2,3], Yu-Ao Chen[1,2,3], & Jian-Wei Pan[1,2,3,4]

[1]*Shanghai Branch, National Laboratory for Physical Sciences at Microscale and Department of Modern Physics, University of Science and Technology of China, Hefei, Anhui 230026, China*

[2]*CAS Center for Excellence and Synergetic Innovation Center in Quantum Information and Quantum Physics, University of Science and Technology of China, Shanghai, 201315, China*

[3]*CAS-Alibaba Quantum Computing Laboratory, Shanghai, 201315, China*

[4]*Physikalisches Institut, Ruprecht-Karls-Universität Heidelberg, Im Neuenheimer Feld 226, 69120 Heidelberg, Germany*

**Abstract**: We report a new apparatus for the study of two-species quantum degenerate mixture of $^{41}$K and $^6$Li atoms. We develop and combine several advanced cooling techniques to achieve both large atom number and high phase space density of the two-species atom clouds. Furthermore, we build a high-efficiency two-species magnetic transport system to transfer atom clouds from the 3D magneto-optical-trap chamber to a full glass science chamber of extreme high vacuum environment and good optical access. We perform a forced radio-frequency evaporative cooling for $^{41}$K atoms while the $^6$Li atoms are sympathetically cooled in an optically-plugged magnetic trap. Finally, we achieve the simultaneous quantum degeneracy for the $^{41}$K and $^6$Li atoms. The Bose-Einstein condensate of $^{41}$K has $1.4 \times 10^5$ atoms with a condensate fraction of about 62%, while the degenerate Fermi gas of $^6$Li has a total atom number of $5.4 \times 10^5$ at 0.25 Fermi temperature.

## 1. Introduction

Ultracold mixture of dilute gases is a flourishing research field in recent years. With highly controllable experimental parameters, it serves as an ideal platform to study few- and many-body quantum phenomena occurring in various systems, such as condensed matter physics, nuclear physics, and astrophysics [1]. Of particular interest is the ultracold mixture of Bose and Fermi gases. It may provide rich insights into, for example, polaron physics [2, 3], quantum chromodynamics matter [4], and theoretical models of High-Tc superconductivity [5]. Moreover, mixture of different atomic species could introduce large mass imbalance which significantly changes the inter-species interactions. Over the past decades, several two-species Bose-Fermi mixtures have been realized [6-9]. The initial study of Bose-Fermi mixtures focus on the realization of degenerate Fermi gas [10, 11], where the bosonic component is served as a coolant. With the development of cooling and trapping techniques, many important quantum phenomena arising from the inter-species interactions have also been demonstrated, such as Efimov states [12-14], Bose polaron [15], and polar molecules [16-18]. Very recently, a mixture of Bose and Fermi superfluids has been realized [19, 20].

In this Article, we report a new apparatus for producing degenerate Bose-Fermi mixture of $^{41}$K and $^6$Li atoms, which have a mass ratio of about 7. We describe in detail the vacuum system and laser setups. We also discuss the laser cooling strategy for both species and present a detailed implementation of it. Particularly, many state-of-the-art laser cooling techniques are employed, such as an advanced 2D$^+$ magneto-optical trap (MOT) [21, 22] for $^{41}$K atoms, a D1 line gray molasses for $^{41}$K atoms [23-28], and an ultraviolet (UV) MOT for $^6$Li atoms [29-31] etc. We introduce a high field D1 optical pumping method [26] to achieve perfect spin purification and high efficiency transfer for the magnetic trapping. Furthermore, we design and construct a magnetic transport system, achieving high efficiency transport for both species.

The experimental procedure starts from generating intense slow atomic beams, which are provided by a spin-flip Zeeman slower for $^6$Li atoms and an advanced 2D$^+$ MOT for $^{41}$K atoms. Then we simultaneously capture and cool the atoms by a two-species 3D MOT. Next, the $^6$Li atoms are further cooled to higher phase space density (PSD) by a compressed-MOT (CMOT) while the $^{41}$K atoms are cooled by a novel D1-D2 CMOT. To achieve lower temperature and higher PSD, D1 line gray molasses for $^{41}$K atoms and UV MOT for $^6$Li atoms are further implemented. The two-species are then prepared to the stretched states by high field D1 optical pumping for magnetic trapping and spin purification. After transferred to the magnetic quadrupole trap, they are adiabatically transported to the science chamber with extreme high vacuum environment, where an optically-plugged magnetic trap is used to confine the two species. Finally, with forced radio-frequency (RF) evaporative cooling on $^{41}$K atoms in the optically-plugged magnetic trap, we achieve the quantum degenerate mixture of $^6$Li and $^{41}$K atoms. The obtained degenerate mixture contains $1.4\times10^5$ $^{41}$K atoms with a condensate fraction of 62% and $5.4\times10^5$ $^6$Li atoms at 0.25 Fermi temperature.

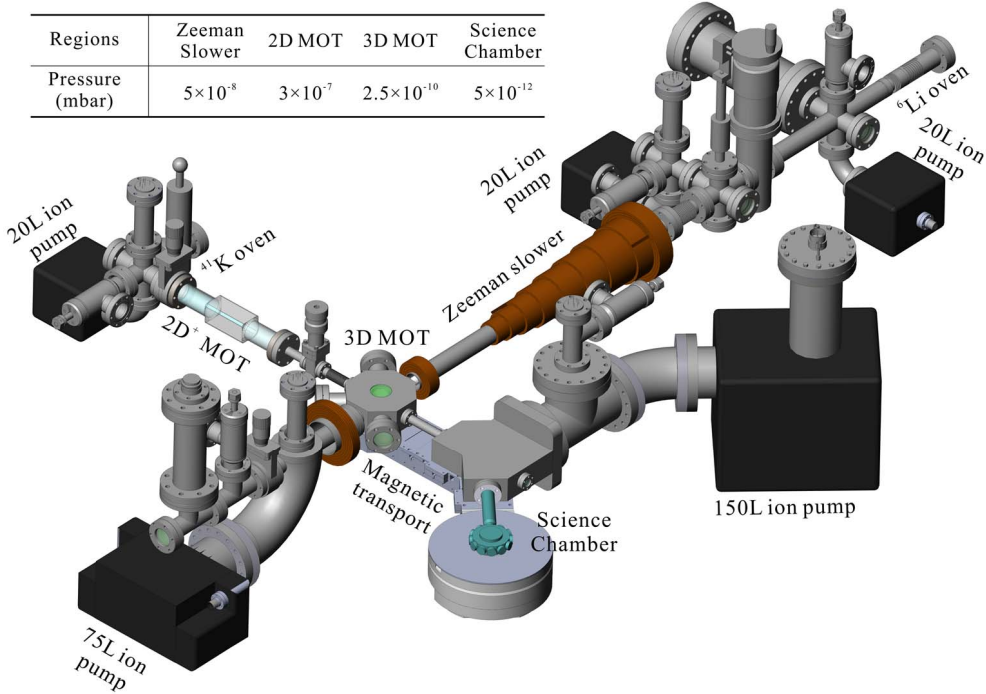

Figure 1. Schematic view of vacuum system.

## 2. Vacuum system

Ultra-high vacuum environment is necessary for ultracold atom experiments. It provides an extreme isolation from the thermal environment, which is essential for the cooling, manipulating, and probing of atoms. Our vacuum system contains three main parts, the lithium vacuum, the potassium vacuum, and the experimental chambers (see figure 1). The lithium vacuum is consisted of a custom-designed tapered-reflux oven and a spin-flip Zeeman slower. The potassium vacuum contains an oven and a $2D^+$ MOT glass cell. For achieving ultra-high vacuum in the experimental chambers, two differential pumping tubes are installed in the lithium and potassium vacuums, having the conductance of 0.03 L/s and 0.06 L/s, respectively. The MOT chamber lies at the center of the vacuum system, connects two atomic vacuums and the science chamber. It is an octagonal chamber with two windows directly welded into the upper and lower surfaces. The science chamber is a dodecagonal full glass cell, allows us to apply different laser configurations. Two high quality and large diameter windows in the vertical direction are used for high-resolution imaging. The two chambers are connected by a 100 mm long differential pumping tube and a special-designed $135°$ transport chamber. By utilizing four ion pumps, two titanium sublimation pumps, and a 150 L combination pump, an extremely high vacuum with background gas pressure of $5\times10^{-12}$ mbar in the science chamber is achieved. The inserted table in figure 1 shows measured vacuum pressures of different regions in the system.

### 3. Laser setups

The realization of two-species degenerate mixture of $^{41}$K and $^6$Li atoms requires exquisite control of complex laser systems. In our experiment, cooling lasers with four different wavelengths are used, i.e., 767 nm for $2D^+$ MOT and 3D MOT of $^{41}$K atoms, 770 nm for D1 line gray molasses of $^{41}$K atoms, 671 nm for Zeeman slower and 3D MOT of $^6$Li atoms, and 323 nm for UV MOT of $^6$Li atoms. A sketch of the energy levels and transitions of interest is shown in figure 2, while the simplified laser setups are shown in figure 3 and figure 4.

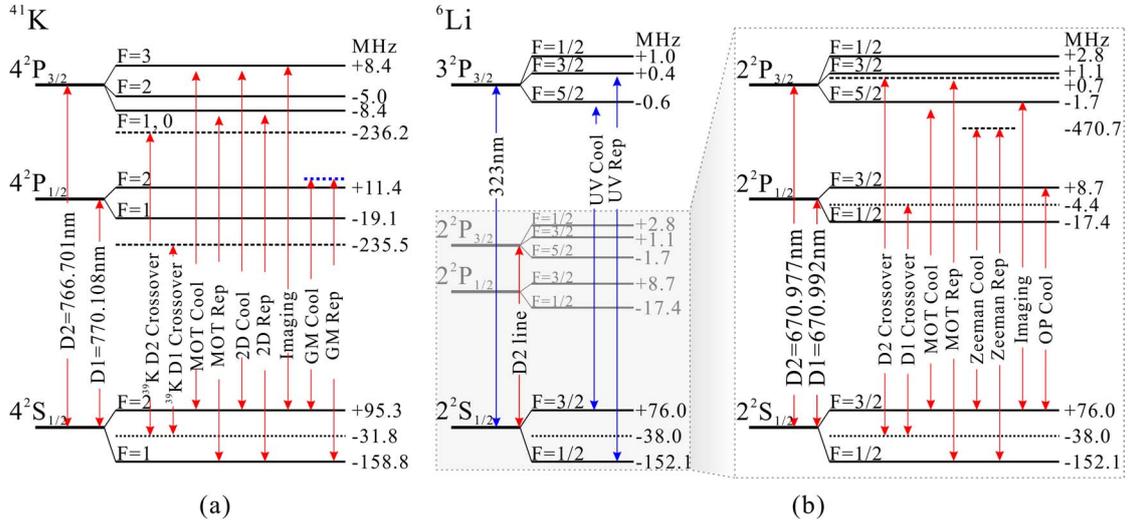

Figure 2. Energy level diagram and transition lines of $^{41}$K and $^6$Li atoms. The black horizontal lines represent the energy levels. The red vertical lines with arrows represent the D1 and D2 transitions required for both species. Blue vertical lines with arrows in the center represent transitions used for the UV MOT. The "GM" in (a) and "OP" in (b) are the abbreviations of "Gray Molasses" and "Optical Pumping", respectively..

The laser setup for manipulating $^{41}$K atoms is depicted in figure 3. Seven diode lasers (DLs)

**3 / 18**

and four tapered amplifiers (TAs) are used. DL1 serves as the master laser with its frequency stabilized to the D2 transition of $^{39}$K atoms via the frequency-modulation (FM) spectroscopy [32] as described below. A small portion of DL1 passes through a single-pass (SP) acoustic-optical modulator (AOM). The first order diffraction laser beam, with a frequency shift of +120 MHz, is then modulated by a 50 MHz electro-optical modulator (EOM) and used for FM spectroscopy. The output signal of the photodiode (PD-K 1) is mixed with the reference signal, yielding the derivative of the saturated absorption spectroscopy. The resulting error signal is fed into a PID controller to control the driving current of DL1 with a loop bandwidth of 100 kHz. All of the DL2—DL6 are frequency-locked to the master laser with specific frequency differences. The frequency stabilization is accomplished by the tunable frequency-offset (FO) locking [33]. By superimposing two laser beams with the same polarization into a fast photodiode and utilizing a voltage-controlled oscillator (VCO), two mixers, and a delay line, we generate the error signal in the form of $sin(|\nu_0 - \nu_{VCO}|\tau)$ ($\nu_0$ is the frequency difference of two lasers and $\nu_{VCO}$ is the output frequency of the VCO, $\tau$ is the time delay). Since the frequency locking requires $\nu_0=\nu_{VCO}+f_c$ where $f_c$ is a constant frequency depends on the length of the delay line, one can dynamically tune the frequency difference by changing the output frequency of the VCO. The outputs of DL2—DL5 are injected into the four TAs (TA1—TA4), respectively. DL6 is fiber coupled and used for imaging. The output of TAs pass through several AOMs for frequency shifting and intensity tuning. Combined with additional mechanical shutters, the AOMs also serve as fast optical switches. All of the lasers are coupled by single mode polarization maintenance (PM) optical fibers and used for the 2D pushing/cooling/repumping, 3D cooling/repumping, and imaging.

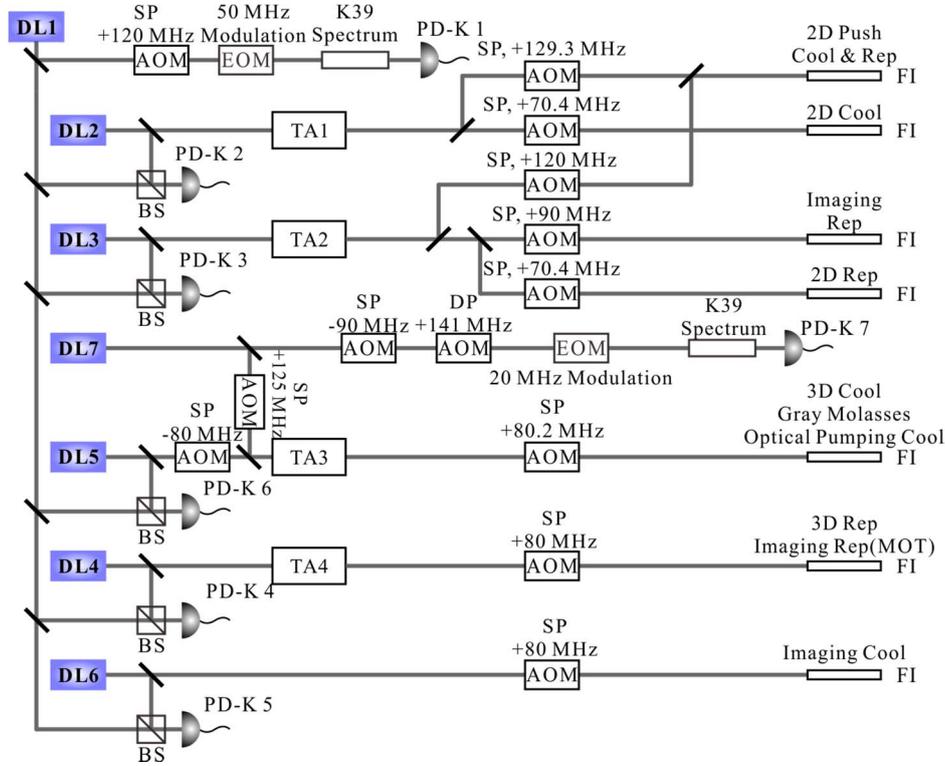

Figure 3. Simplified laser setup for $^{41}$K laser cooling, optical pumping, and imaging.

DL7 is used for the D1 line gray molasses cooling. It is locked to the D1 transition of $^{39}$K

atoms through modulation transfer spectroscopy [34] for achieving better locking stability. To bridge the gap between the D1 transition frequency of $^{41}$K and $^{39}$K atoms, two AOMs, one in SP configuration and the other in double-pass (DP) configuration, are used for the spectroscopy. The DP AOM can also be used to dynamically tune the frequency of DL7 while laser is locked. A resonant EOM (center frequency of 254 MHz) is employed to generate the required repumping frequency for the D1 line gray molasses. The dichromatic laser is superimposed with seed laser of 3D cooling (DL5) through a custom-designed dichroic mirror before injecting into the TA3. The AOMs used in DL5 and DL7 serve as fast switches which allow us to rapidly change the output frequency for different cooling phases.

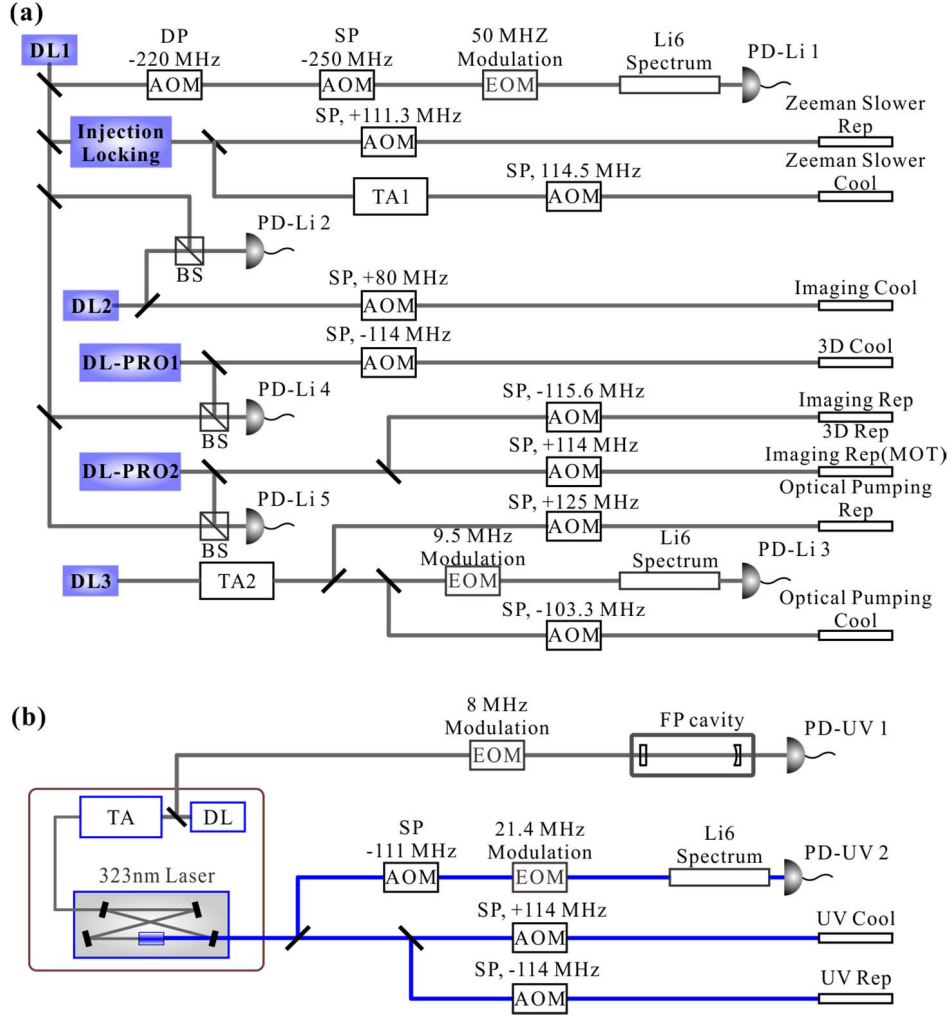

Figure 4. Simplified laser setups for $^6$Li atoms. (a) Laser setup for $^6$Li laser cooling, optical pumping, and imaging. (b) Laser setup for $^6$Li UV MOT.

The laser setup for D2-line cooling, imaging, and optical pumping of $^6$Li atoms is shown in figure 4(a). Similar to the laser setup of $^{41}$K atoms, the frequency of lithium master laser (DL1) is locked to the D2 transition of $^6$Li atoms via FM spectroscopy. The DL2 is used for imaging with the frequency-stabilized to DL1 through tunable FO locking. The DL3 is the seed laser of TA2 which is used for D1 optical pumping. The frequency stabilization of DL3 is demonstrated through modulation transfer spectroscopy of the D1 transition. Two combination laser systems (TA-Pros) are used as the main D2-line cooling lasers of $^6$Li atoms. They are stabilized to DL1 through tunable FO locking. We also build an injection locked laser with about 5 mW laser from DL1 being injected into it. The output of the injection locked laser is split into two part, one is used for the repumping of Zeeman slower, and the other is served as the seed laser of the Zeeman slower cooling TA (TA1). The required frequency shifting, intensity tuning, and optical switching are achieved by a series of AOMs.

Figure 4(b) shows the UV laser setup for narrow line cooling of $^6$Li atoms. The main laser here is a commercial laser with 60 mW output power and 1 MHz linewidth at 323 nm. Considering the UV transition ($2S_{1/2} \rightarrow 3P_{3/2}$) of $^6$Li is rather narrow, we have to reduce the linewidth of the laser which is achieved by locking it to a homemade high fineness Fabry-Perot cavity. To stabilize the cavity to the UV transition, we perform a UV FM spectroscopy of $^6$Li atoms by using a small portion of the UV laser. With these techniques, we eventually achieve both frequency stabilization and narrow linewidth (30 kHz) of the UV laser. The main output of the UV laser is divided into two laser beams. They pass through two SP AOMs for shifting the frequency of +114 MHz (-114 MHz) respectively, serving as the cooling and repumping lasers for the UV MOT.

### 4. Laser cooling and trapping

Simultaneous cooling and trapping of two-species atomic mixture is challenging. Tremendous efforts are required for achieving both large atom number and high PSD. In this section, we give a detail description of our sophisticated laser cooling and trapping strategy, including a spin-flip Zeeman slower [35] for $^6$Li atoms, an advanced 2D$^+$ MOT for $^{41}$K atoms, a two-species 3D MOT and CMOT, a D1 line gray molasses for $^{41}$K atoms, and a UV MOT for $^6$Li atoms.

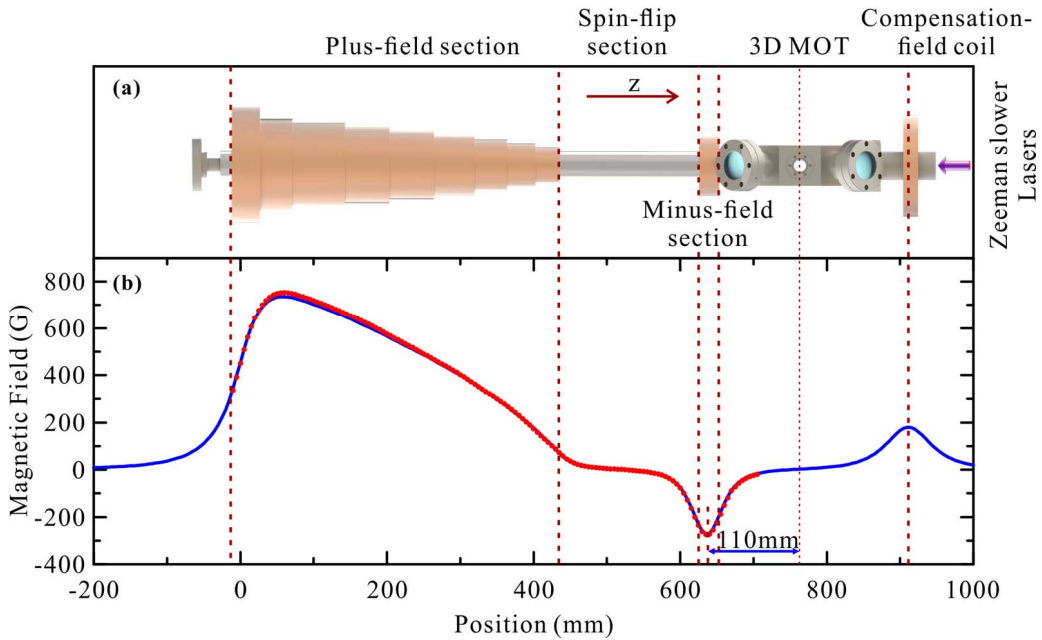

Figure 5. The schematic view (a) and the calculated (blue line) and measured magnetic field (red dot) of Zeeman slower (b).

*4.1. Spin-flip Zeeman slower for $^6$Li*

The lithium atomic beam is produced by the oven which consists of a heating cell and a tapered-reflux nozzle. The heating cell is heated to 450 ℃ to generate the required vapor pressure of

⁶Li atoms. The tapered reflux nozzle collects the blocked atoms, extending the lifetime of the oven. A 150 mm long tube with 3.5mm inner diameter is installed at the end of the reflux nozzle, serving as differential pumping tube and atomic beam collimator. After collimating, the transverse divergence of the atomic beam is reduced to 10 mrad. However, the longitudinal velocity of lithium atoms is on the order of 1000 m/s, far beyond the MOT's capture range. We adopt the standard Zeeman slowing technique to decelerate the longitudinal velocity of the ⁶Li atoms. By creating an inhomogeneous magnetic field, the frequency shift accumulated during the deceleration is compensated by the Zeeman effect, leading to a continuous deceleration.

The configuration of the spin-flip Zeeman slower is shown in figure 5(a). The magnetic coil system consists of three sections, i.e., plus-field section, spin-flip section, and minus-field section. The plus-field section is a cone-shaped coil of 11 layers, having a length of 437.4 mm. The minus-field section is a concentric coil of 4×4 windings. The spin-flip section connects the plus- and minus-field sections with a length of 185.6 mm. The coil assembly extends over 650 mm and the minimum point of magnetic field is 110 mm away from the center of MOT chamber. To compensate the residual magnetic field created by the minus-field section, we install a concentric coil at the opposite position of the MOT chamber (see figure 5 (a)). In the experiment, the currents for the plus and minus coils are optimized to be 45 A and 62 A, while the current for the compensation coil is set to 23 A. The calculated and measured magnetic field of Zeeman slower are shown in figure 5(b), having a perfect agreement. The resulting magnetic field in the MOT center is calculated to be -0.03 G with a gradient of 2.16 G/cm, ignorable for 3D MOT.

Slowing and repumping lasers are superimposed by a beam splitter and expanded by a telescope. Then the dichromatic laser beam is focused by a 500 mm achromatic lens. The $1/e^2$ beam diameter at the center of 3D MOT is about 16 mm. The peak intensity of slowing light is 15.67 $I_{sat}^{Li}$ ($I_{sat}^{Li}$=2.54 mW/cm² is the saturation intensity of ⁶Li), while that of repumping light is 5.49 $I_{sat}^{Li}$. Both the slowing and repumping lasers are $\sigma^+$ polarized and the optimal detuning is -80.07 $\Gamma_{Li}$ ($\Gamma_{Li}$=2π×5.87 MHz is the D2 transition linewidth of ⁶Li). With such a Zeeman slower, we can obtain an atomic beam with $2\times10^{10}$ atoms/s flux and about 20 m/s average velocity.

*4.2. Advanced 2D⁺ MOT for ⁴¹K*

2D MOT is an efficient pre-cooling method to produce atomic beams with high flux and low velocity. However, in the conventional 2D MOT, cooling and trapping occur only in the transverse directions which makes cooling of atoms inefficient. In the case ⁴¹K atoms whose natural abundance is 6.73%, the loading rate is limited. To improve the performance of ⁴¹K pre-cooling stage, we adopt an advanced 2D⁺ MOT scheme in the experiment. The key ingredient is to applying a pair of retro-reflect cooling and repumping beams in the axial direction, leading to an axial optical molasses cooling. The atoms which are axially cooled will experience more transverse cooling time, resulting in less transverse divergence and higher density of the atomic beams.

The schematic view of our advanced 2D⁺ MOT is shown in figure 6. The 2D⁺ MOT chamber is a cuboid full glass cell with a size of 45×45×95 mm. A custom-designed differential pumping tube is connected to the glass cell, onto which a mirror and an absorptive polarizer are glued. Both the polarizer and the mirror are drilled a hole in the center with a diameter of 2 mm. We wind two pairs of rectangular coils around the glass cell to generate the required cylindrical quadrupole magnetic field in the transverse directions, while the magnetic field along the axial direction is zero. Two additional coil pairs in Helmholtz configuration are used to shift the center of the 2D$^+$ MOT.

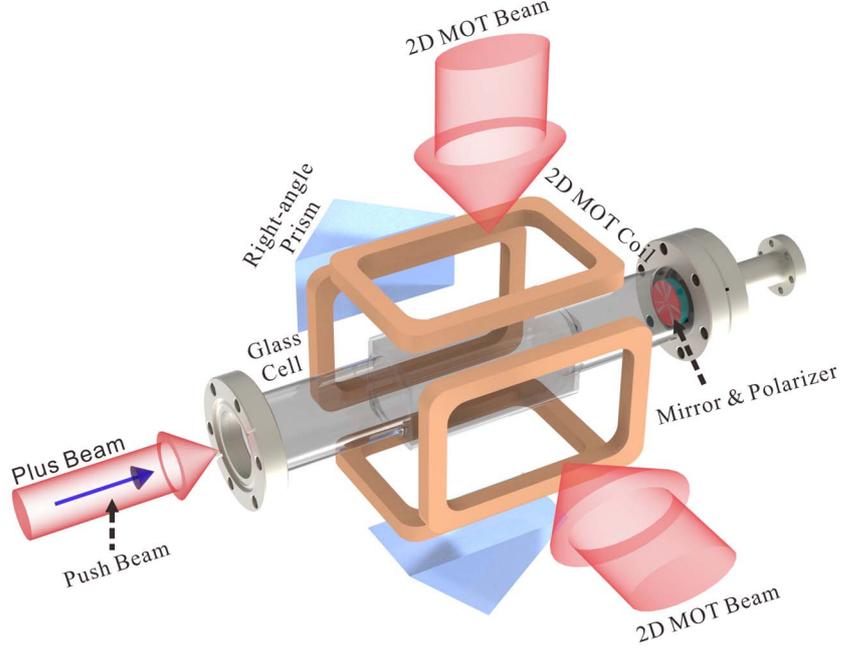

Figure 6. Schematic view of the advanced 2D$^+$ MOT.

The cooling and repumping lasers are superimposed by a polarizing beam splitter (PBS). The resulting dichromatic laser beam is separated into 3 beams and expanded by spherical and/or cylindrical telescopes to create the transverse and axial beams. Two transverse beams have an elliptical cross section (1/e$^2$ diameters: 30 mm and 90 mm). They are σ$^+$ polarized and retro-reflected by two custom designed right-angle prisms with phase compensation coatings. The axial laser beam is linearly polarized with a 1/e$^2$ diameter of 20 mm. It passes through the absorptive polarizer and then retro-reflected by the mirror. The intensity ratio between the inward and outward beams can be freely tuned by a half wave plate (HWP). Additionally, an independent blue detuned push beam is applied to increase probability of atoms passing through the differential pumping tube and reaching the 3D MOT chamber. The push beam contains both the cooling and repumping frequencies with a 1/e$^2$ diameter of 1.2 mm, which can fully pass through the 2 mm hole. Finally, with the optimization of magnetic field gradient (11.2 G/cm) and other parameters (see table 1), a maximum loading rate of $2.2 \times 10^9$ atoms/s for the single species $^{41}$K 3D MOT is achieved, which is more than one order of magnitude larger in comparison with the conventional 2D MOT [36, 37].

Table 1. Optimal parameters of the advanced 2D$^+$ MOT

|  | Transverse cooling | Axial cooling | Push |
|---|---|---|---|
| $I_{cool}$ ($I_{sat}^K$ =1.75 mW/cm$^2$) | 5 | 30.77 | 50.55 |
| $I_{rep}$ ($I_{sat}^K$) | 5 | 30.77 | 50.55 |
| $\Delta\omega_{cool}$($\Gamma_K$=2π×6.04 MHz) | -3.64 | -3.64 | +6.11 |
| $\Delta\omega_{cool}$($\Gamma_K$) | -2.15 | -2.15 | +6.06 |

## 4.3. Two species 3D MOT and Compressed MOT

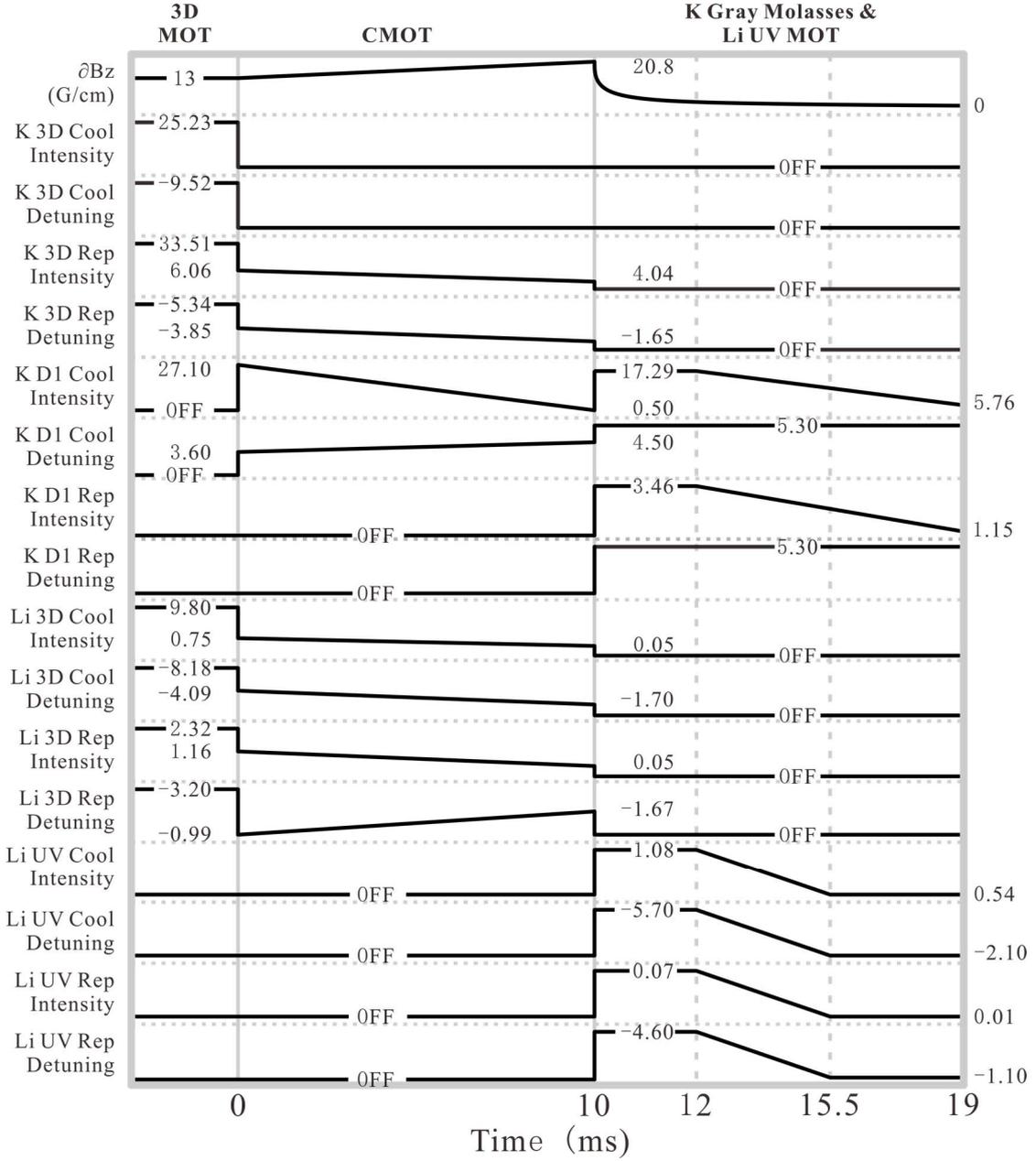

Figure 7. Time sequence and experimental parameters for two-species laser cooling. The intensity of K laser, Li laser, and UV laser are in $I_{sat}^{K}$, $I_{sat}^{Li}$, and $I_{sat}^{uv}$, respectively. The detuning of K laser, Li laser, and UV laser are in $\Gamma_{K}$, $\Gamma_{Li}$, and $\Gamma_{3P}$, respectively.

3D MOT is a standard technique to obtain cold atom clouds. It consists of six counter-propagating laser beams and a quadrupole magnetic field. The magnetic field gradient for loading the two species is optimized to be 13 G/cm. For simultaneously manipulating the $^6$Li and $^{41}$K lasers, broadband PBS, dual-wavelength HWPs, and quarter-wave plate (QWPs) are used. Each 3D MOT laser beam contains four frequencies, i.e. cooling and repumping frequencies of $^{41}$K and $^6$Li atoms. The $1/e^2$ beam diameter is 18 mm, other parameters such as the laser intensity and detuning are shown in figure 7. We mention that due to the unresolved excited hyperfine structure, there's no cycling transition for both $^{41}$K and $^6$Li atoms. Therefore, the repumping laser also contribute to the cooling process that requires large intensity. Figure 7 shows the optimized loading parameters for the two-species 3D MOT. With a 2 s MOT loading phase, a total number of $4\times10^9$ $^{41}$K atoms and $3\times10^9$ $^6$Li atoms can be obtained simultaneously. The temperature of $^{41}$K atoms is 5.6 mK while that of $^6$Li atoms is 2.9 mK.

To increase the PSD, we perform two different CMOT for $^{41}$K and $^6$Li atoms simultaneously. During the CMOT phase, the magnetic field gradient is linearly ramped from 13 G/cm to 20.8 G/cm in 10 ms. For $^6$Li atoms, the peak intensity of 3D cooling and repumping laser beams are suddenly set to 0.75 $I_{sat}^{Li}$ and 1.16 $I_{sat}^{Li}$, respectively. Then both of them are linearly ramped to 0.05 $I_{sat}^{Li}$ in 10 ms. Similarly, the frequency of cooling and repumping light are suddenly shifted to -4.09 $\Gamma_{Li}$ and -0.99 $\Gamma_{Li}$, respectively. Then they are linearly ramped to -1.70 $\Gamma_{Li}$ and -1.67 $\Gamma_{Li}$ in 10 ms, respectively (see figure 7). After the CMOT, the temperature of $^6$Li atoms reduces to 284 μK with an atom number of $2.9\times10^9$, resulting a PSD of $5.2\times10^{-7}$. For $^{41}$K atoms, a novel D1-D2 CMOT technique [27] is adopted. Comparing with the $^6$Li CMOT, the D1 cooling transition |F=2⟩ to |F'=2⟩ of $^{41}$K is used instead of the D2 cooling transition. In order to suppress the light-assisted collisions, cooling (repumping) peak intensity is linearly reduced from 27.10 $I_{sat}^{K}$ (6.06 $I_{sat}^{K}$) to 0.50 $I_{sat}^{K}$ (4.04 $I_{sat}^{K}$) while the detuning is ramped from +3.6 $\Gamma_K$ (-3.85 $\Gamma_K$) to +4.5 $\Gamma_K$ (-1.65 $\Gamma_K$) in 10 ms, respectively. With this technique, the temperature of $^{41}$K atoms is reduced to 247 μK. The atom number of D1-D2 CMOT is $2.6\times10^9$ and PSD is $1.13\times10^{-7}$.

*4.4. $^{41}$K D1 line gray molasses and $^6$Li UV MOT*

In a MOT, the lowest achievable temperature is Doppler temperature $T_D = \hbar\Gamma/(2k_B)$ (where $\hbar$ is the Plank constant divided by $2\pi$, $\Gamma$ is the linewidth of transition, and $k_B$ is the Boltzmann constant). Sub-Doppler cooling well below $T_D$ is generally achieved with Sisyphus cooling in optical molasses. However, due to the unresolved excited-state hyperfine structure of the D2 transition, standard molasses cooling of $^6$Li and $^{41}$K are inefficient. In order to further increase the PSD, we adopt an advanced cooling scheme, where sub-Doppler cooling in gray molasses for $^{41}$K atoms [27] and UV MOT for $^6$Li atoms are employed, respectively. In order to combine the two different laser cooling technique, tremendous efforts have been devoted to optimize the relevant experimental parameters (see figure 7). As more $^{41}$K atoms are required, we suddenly switch off the magnetic gradient field after the CMOT phase to perform the gray molasses of $^{41}$K. The UV MOT of $^6$Li is applied at the residual magnetic field.

D1 line gray molasses is a newly developed laser cooling technique based on the combination of D1 line Sisyphus cooling and velocity-selective-coherent-population-trapping (VSCPT) in a Λ-type three-level system. In order to implement gray molasses, cooling and repumping lasers with frequencies blue-detuned from the $^{41}$K D1 transition line are required (see figure 2). As described in the third section, the same optical setup is used for both D1 line gray molasses and D2 MOT, which greatly reduces the experimental complexity. At the beginning of the gray molasses, the D2 repumping laser and magnetic field are switched off while the input signal of EOM is turned on, generating the required cooling and repumping frequencies. Experimentally, we use the same detuning of +5.3 $\Gamma_K$ for the two lasers to fulfill the Raman condition and the cooling/repumping intensity ratio is fixed at about 5:1. To achieve high capture efficiency of the gray molasses, maximum laser intensity is applied in the first 2 ms, which is 17.29 $I_{sat}^{K}$ and 3.46 $I_{sat}^{K}$ for the cooling and repumping light, respectively. Then we simultaneously ramp the laser intensity down to 5.76 $I_{sat}^{K}$ (1.15 $I_{sat}^{K}$) in the following 7 ms. With the gray molasses cooling, the temperature of $^{41}$K atoms is reduced to 42 μK and the PSD is increased to $5.4\times10^{-6}$, providing an excellent starting point for further magnetic trapping and evaporative cooling.

The motivation for implementing a $^6$Li UV MOT is simple: the linewidth of UV transition for $^6$Li is much narrower than that of D2 transition, lead to a lower Doppler temperature. For example, the overall linewidth of the 2s-3p transition of $^6$Li is $\Gamma_{3P}=2\pi\times754$ kHz, corresponding to a Doppler temperature of 18 μK which is much lower than that of D2 transition (141 μK). In the experiment, the UV MOT is performed in parallel with the gray molasses of $^{41}$K (see figure 7). The cooling and repumping lasers are superimposed by a PBS and then divided into six balanced laser beams by a series of PBSs and HWPs. The UV laser beams are superimposed with the red MOT laser beams through custom designed dichroic mirrors, where the 323 nm lasers are transmitted while the 671 nm, 767 nm, and 770nm are reflected. The $1/e^2$ beam diameter is 9.4 mm with maximum intensity of 1.15 $I_{sat}^{uv}$ (where $I_{sat}^{uv}$ =13.8 mW/cm$^2$ is the saturation intensity of $^6$Li UV transition). The weak magnetic confinement is provided by the residual magnetic field gradient after switching off the current of the MOT coils. At the first 2 ms of the UV MOT, maximum intensity and largest detuning of the cooling (repumping) lasers are used to collect as many atoms as possible. Next, we ramp down the cooling (repumping) intensity of each beam from 1.08 $I_{sat}^{uv}$ (0.07 $I_{sat}^{uv}$) to 0.54 $I_{sat}^{uv}$ (0.01 $I_{sat}^{uv}$) in 3.5 ms. The cooling (repumping) detuning are decreased from -5.70 $\Gamma_{3P}$ (-4.70 $\Gamma_{3P}$) to -2.10 $\Gamma_{3P}$ (-1.10 $\Gamma_{3P}$) at the same time. This step is analogue with a red CMOT, where the light-assisted collisions are suppressed. By holding these parameters for another 3.5 ms, a cloud of $1.24\times10^9$ atoms with a temperature of 62 μK is obtained. The calculated PSD is $5.5\times10^{-5}$, which is two orders of magnitude larger than that of CMOT.

## 5. Optical pumping, magnetic transport, and optically-plugged quadrupole trap

After the laser cooling phase, both of the clouds are magnetically confined and transported to the final glass cell. Then, the atoms are evaporative cooled inside an optically-plugged magnetic trap to achieve the quantum degeneracy. In this section, we will introduce the D1 optical pumping [26], two-species magnetic transport [38], and optically-plugged quadrupole trap.

*5.1. D1 optical pumping*

Magnetic trap is state selective, in which only the atoms at low-field-seeking states can be captured. However, the atoms are depolarized after the laser cooling phase as they stay in both hyperfine states (F states) and distribute randomly in all the Zeeman states ($m_F$ states). Therefore, atoms which occupy high field seeking states or states which are susceptible to undergo spin relaxation will loss from the magnetic trap. Moreover, heating and atom loss induced by the interspecies spin-exchange collisions will significantly reduce the lifetime of atoms in the magnetic trap. Therefore, it's necessary to pump all the atoms into the stretched states before the magnetic trap is switched on.

In the experiment, we implement a high field D1 optical pumping for both species to achieve high efficiency transfer and perfect spin purification. The target states are |F=2, m$_F$=2⟩ of $^{41}$K and |F=3/2, m$_F$=3/2⟩ of $^6$Li, respectively, as there's no channel for interspecies spin-exchange collision and both of the states have the strongest confinement in the magnetic trap. The advantage of D1 optical pumping is that the target states are dark states at the same time.

For instance, if we drive |F=2⟩ to |F'=2⟩ transition of $^{41}$K, the atoms which are pumped into the $m_F$ = 2 state will not absorb any photons and thus they are protected against further heating.

Table 2. Optimal parameters for D1 optical pumping

|  | $^{41}$K Optical Pumping | $^{6}$Li Optical Pumping |
|---|---|---|
| $I_{cool}$ ($I_{sat}^{K/Li}$) | 1.3 | 0.5 |
| $I_{rep}$ ($I_{sat}^{K/Li}$) | 35.9 | 3.9 |
| $\Delta\omega_{cool}$ ($\Gamma_{K/Li}$) | +5.3 | -0.68 |
| $\Delta\omega_{rep}$ ($\Gamma_{K/Li}$) | -0.1 | -2.38 |

At the end of laser cooling phase, a homogeneous magnetic field of 15 G is turned on in 0.75 ms, providing large separation of different Zeeman states which is essential for the spin purification. Then, a pair of balanced counter-propagating light beams containing D1 pumping and repumping frequencies with a duration of 50 us are applied to both species. Both of the pumping and repumping lasers are $\sigma^+$ polarized. With optimized parameters (see table 2), we obtain 70% and 90% pumping efficiency for $^{41}$K and $^{6}$Li atoms, respectively. By using RF spectroscopy, we also verify that nearly all the atoms are populated in the target state.

*5.2. Two-species magnetic transport*

After loading both the species into the magnetic trap, they are magnetically transported to a glass cell with an extreme high vacuum environment and good optical accesses. The principle of magnetic transport is create a moving quadrupole potential which adiabatically displace the magnetic field center and thus the position of atom cloud. This can be done by installing the anti-Helmholtz coils in a movable stage or by applying time-varying currents in a series of overlapping fixed coils [38]. Although the first method is simple, it introduces unwanted noises such as mechanical turbulence and Foucault currents created by the movable stage. In the experiment, we adopt the latter one for the magnetic transport.

During the transport phase, one should keep the aspect ratio of magnetic field gradient in the *x* and *y* direction (see figure 8) as a constant, since the change of aspect ratio will heat the atom clouds. To achieve this goal, a 3-coil transport scheme is implemented as shown in figure 8. At the beginning (see figure 8(a) left), currents run through the first two coil pairs. The atom cloud is confined in the center of the first two coil pairs with an elongated shape. Next, the current is decreased in the first coil pair and simultaneously increased in the third coil pair so that the cloud moves to the center of the second coil pair with the shape unchanged (see figure 8(a) middle). Finally, as the current in the first coil pair decreases to zero and the current in the third coil pair reaches its maximum value, the cloud smoothly moves to the center of the last two coil pairs with a constant shape (see figure 8(a) right). However, due to the large distance between the MOT coils and the first transport coil pair, the aspect ratio of the magnetic field gradient is very large at the beginning of the transport phase. The same situation happens at the end of transport, where the overlap of the quadrupole coils and last transport coil pair is limited. These problems can be solved by using an additional push coil or a pair of push coils which provides an additional magnetic field gradient along the *y* direction to control the aspect ratio.

A sketch of the coil assembly is shown in figure 8(b), including one push coil, one pair of MOT coils, 12 pairs of transport coils, one pair of quadrupole coils, and one pair of final push coils. The atom clouds are transported over a total distance of 537 mm with an angle of 135°, providing an additional optical access along the final transport direction and large operating space around the science chamber. In the first transport section (section 1), the magnetic field gradient in *z* direction is 200 G/cm, then it is gradually decreased to 120 G/cm in the section 2. By using a series of Insulated Gate Bipolar Translators (IGBTs) and PID control loops, controllable time-varying currents running in the coils are generated. With optimized parameters, such as currents, mechanical position of the coils and transport time etc., we can obtain an overall efficiency of about 47% (51%) for $^{41}$K ($^6$Li) atoms in a 3 s transport process. Considering the collisional loss during the transport and loss of atoms passing through the small aperture differential pumping tube (24%), the resulted two-species transport efficiencies are rather high.

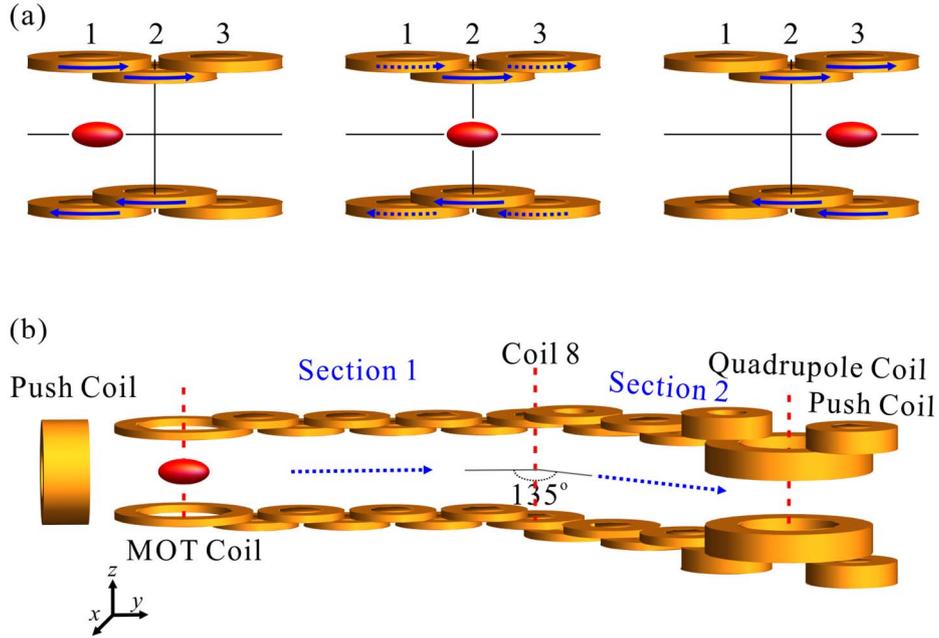

Figure 8. Scheme (a) and configuration (b) of magnetic transport.

*5.3. Optically- plugged quadrupole trap*

Magnetic quadrupole trap is widely used in the ultracold atom experiment as it provides both large trapping volume and strong confinement. However, at the zero point of the magnetic field, Majorana spin-flip [39] occurs which lead to atom loss, heating, and thus a limited lifetime of the atom clouds. The spin-flip region can be estimated by a simple model [40]. By substituting the experimental parameters for the two species, the estimated Majorana spin-flip radius are 2.2 um for $^6$Li atoms and 1.5 um for $^{41}$K atoms, respectively. To prevent this loss mechanism, we apply an optical plug as a repulsive barrier that repels the atoms from the spin-flip region.

In the experiment, we use 532 nm light as the optical plug, which is far blue detuned for both $^{41}$K and $^6$Li atoms. The 532 nm light is provided by a single frequency high power solid laser. To achieve better beam profile and stability, high power fiber coupling is implemented. Finally, an 11 W laser beam is tightly focused to the center of the atom clouds with a $1/e^2$ beam diameter of 38 μm. The resulting repulsive barriers for $^{41}$K and $^6$Li atoms are 1.15 mK and 1.19 mK, respectively. The alignment of the focused beam position is very critical, which is achieved by a pico-motor based mirror mount. With the presence of optical plug, the measured lifetime of atom clouds in the magnetic trap are more than 1 minute.

## 6. Mixture of degenerate $^{41}$K and $^6$Li atom clouds

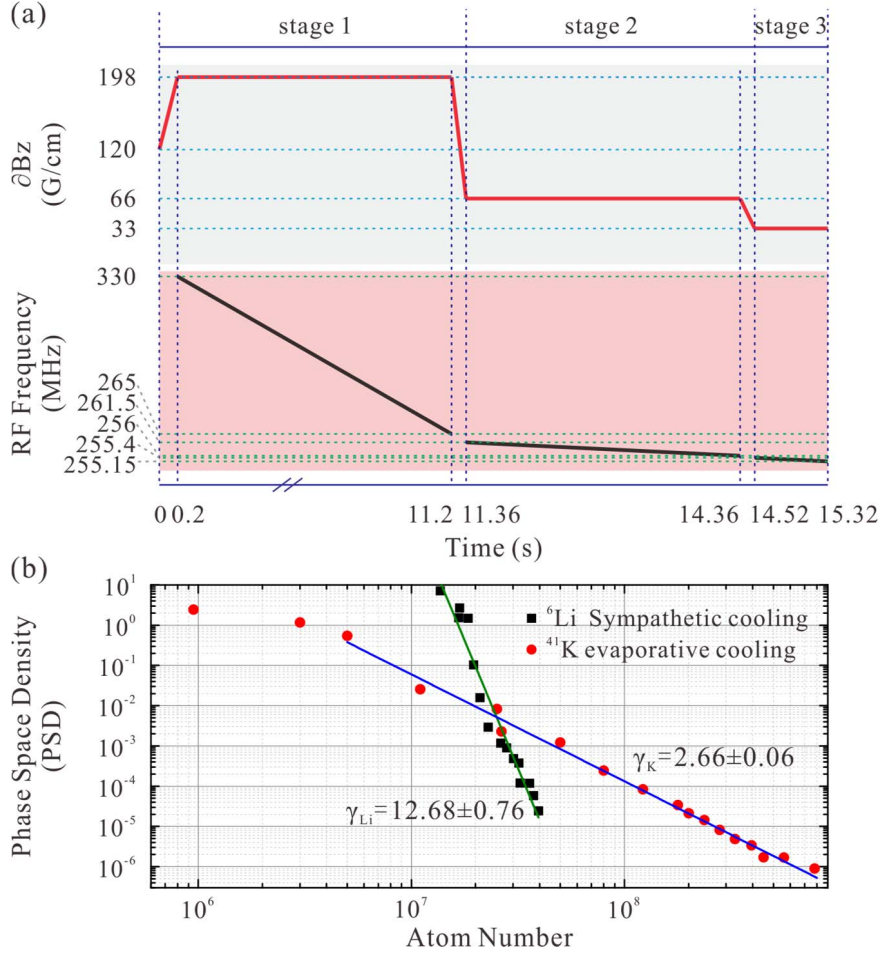

Figure 9. Evaporation time sequence (a) and evolution of the PSD with atom number (b).

Evaporative cooling [41] is required to achieve two-species quantum degeneracy, which can be performed in the optically-plugged magnetic trap with long lifetime and tight confinement. However, s-wave collisions of single-spin $^6$Li atoms are prohibited due to the Pauli Exclusion Principle. Thanks to the favorable background scattering length between $^{41}$K and $^6$Li atoms [42], we can implement an efficient sympathetic cooling scheme to solve this problem. In the experiment, we apply the forced RF evaporative cooling to $^{41}$K atoms by driving $|F = 2, m_F = 2\rangle \rightarrow |F = 1, m_F = 1\rangle$ transition, while the $^6$Li atoms are cooled sympathetically. During this process, the consumption of $^6$Li atoms is much less than that of $^{41}$K atoms. Therefore, we only load 100 ms $^6$Li atoms in the 3D MOT phase while the $^{41}$K atoms are maximized. Finally, a lithium cloud of $4\times10^7$ atoms with a PSD of $2.45\times10^{-5}$ and a potassium cloud of $7.8\times10^7$ atoms with a PSD of $1\times10^{-6}$ are simultaneous obtained in the optically-plugged magnetic trap.

The evaporative cooling process is carefully optimized and can be divided into three stages (see figure 9(a)). In the first 200 ms of stage 1, the magnetic field gradient is ramped to 198 G/cm to enhance the elastic collision rate in the trap. Then we linearly sweep the RF signal from 330 MHz to 265 MHz in 11 s. At the end of this stage, the peak density of $^{41}$K atoms is about $3.5\times10^{13}$ atoms/cm$^3$, further evaporation is inefficient due to the increased three-body inelastic collisions. Therefore, we switch off the RF signal, and then ramp down the magnetic field gradient in 160 ms to decompress the trap and lower the peak density of the atom clouds. However, the trap center shifts a lot during the decompress process due to the imperfection of the magnetic coils, leading to the mismatch between the plug beam center and the magnetic field center. To solve this problem, we wind three pairs of coils to create homogenous bias fields in *x*, *y*, and *z* directions. The intensity of the bias fields are stabilized and can be dynamically controlled through a PID system, achieving perfectly overlapping of the trap center with the plug beam center. The second stage evaporation is performed at the gradient of 66 G/cm with RF signal swept from 261.5 MHz to 256 MHz in 3 s. To achieve better quantum degeneracy, the trap is further decompressed by ramping the magnetic field gradient down to 33 G/cm in 160 ms. However, the size of atom clouds are comparable with the beam size of the optical-plug at this stage. Thus the clouds are split into two parts due to the presence of two potential minima created by the optical plug. To create a single potential minimum, we carefully adjust the intensity of horizontal bias field in the perpendicular direction of the optical-plug beam. The center of the magnetic trap is displaced by about 10 μm, large enough for ensuring the existence of only one potential minimum, which is very important for the achieving of the quantum degeneracy. Finally, by sweeping RF signal from 255.4 to 255.15 MHz in 800 ms, a quantum degeneracy mixture with $1.4 \times 10^5$ $^{41}$K atoms and $5.5 \times 10^5$ $^6$Li atoms are obtained (see figure 10). The final hybrid trap is an approximate harmonic trap formed by the optical-plug and magnetic quadrupole potentials. The trap frequencies of $^{41}$K ($^6$Li) are $\omega_x = 479\ (1252)$ Hz, $\omega_y = 140\ (364)$ Hz, and $\omega_z = 252\ (669)$ Hz, where the *x*, *y* are the perpendicular and parallel directions with respect to the optical-plug, and *z* is the vertical direction, respectively.

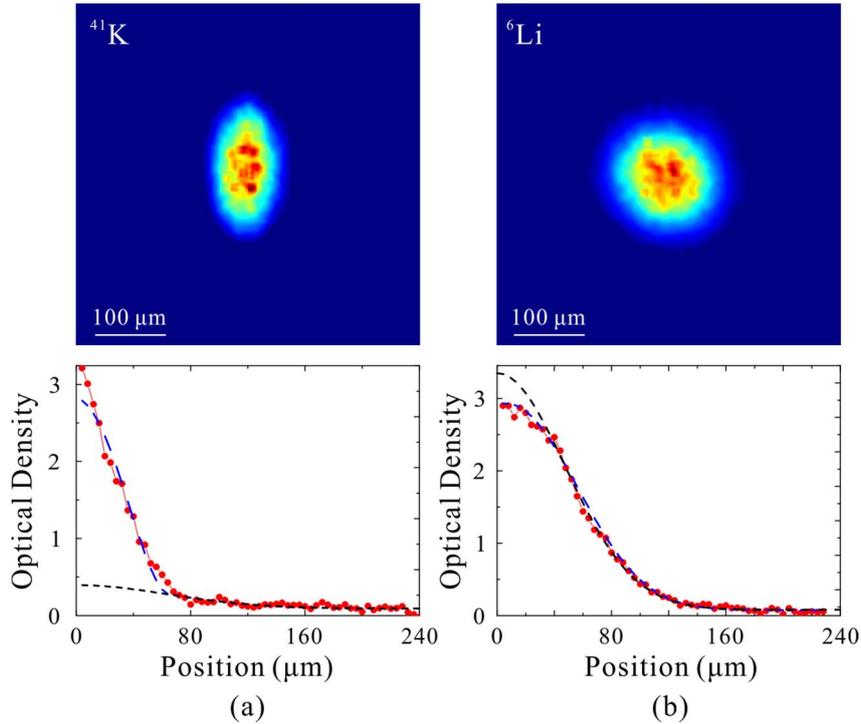

Figure 10. Bose Einstein condensate of $^{41}$K atoms after 15 ms TOF (a) and degenerate Fermi gas of $^6$Li atoms after 1 ms TOF (b). The red dots are the column densities of $^{41}$K and $^6$Li. The blue dash lines are the bimodal distribution fitting for $^{41}$K and Fermi-Dirac distribution fitting for $^6$Li of the column densities, respectively. The dark dash lines are the gauss fitting of the wings of the column densities.

To verify the whole process, we measure the evaporation efficiency $\gamma \equiv -\partial \ln (\text{PSD})/\partial \ln (N)$ for the two-species (see figure 9(b)). The overall $\gamma$ value of the $^{41}$K atoms is $2.66\pm0.06$ while that of $^6$Li atoms is $12.68\pm0.76$, showing a good efficiency of the evaporative cooling process. After evaporative cooling phase, the magnetic trap and the optical-plug laser are switched off simultaneously. Then time-of-flight image are taken after free expansion of 1 ms for $^6$Li atoms and 15 ms for $^{41}$K atoms, respectively. Through a bimodal distribution fitting of the $^{41}$K TOF picture, a condensate fraction of more than 62% is obtained (see figure 10(a)). From the Fermi-Dirac fitting, the temperature of $^6$Li atoms is about 0.25 $T_F$, where $T_F$ is the Fermi temperature (see figure 10(b)). Furthermore, assuming the final trap is approximately harmonic and the two-species are in thermal equilibrium, we can calculate the temperature of $^6$Li atoms by substituting the temperature of $^{41}$K atoms, trapping frequencies, and $^6$Li atom number, resulting in a temperature of about 0.2 $T_F$. Both results are satisfactory and demonstrate the quantum degeneracy of the two species.

## 7. Conclusion

In conclusion, we present an advanced apparatus for studying Bose-Fermi degenerate mixture of $^{41}$K and $^6$Li atoms. We build a complex vacuum system which allows us to independently generate and optimize the two-species atomic beams. More importantly, the vacuum system has a full glass science chamber, providing good optical access and extreme high vacuum environment ($5\times10^{-12}$ mbar) that is essential for the cutting-edge experimental study of ultracold atoms. To achieve fully independent tuning of all the experimental parameters, we build a sophisticated laser setup for simultaneous cooling, manipulating, and probing $^{41}$K and $^6$Li atoms. We adopt several advanced cooling strategies to achieve both large atom number and high phase space density for the two species, such as a spin-flip Zeeman slower, an advanced 2D$^+$ MOT, a D1-D2 compressed-MOT, a D1 line gray molasses, a UV MOT etc. With these laser cooling techniques, $1.24\times10^9$ $^6$Li atoms of $5.5\times10^{-5}$ PSD and $2.4\times10^9$ $^{41}$K atoms of $5.4\times10^{-6}$ PSD are achieved simultaneously. By using a novel high field D1-D1 optical pumping technique, we can transfer 70% $^{41}$K and 90% $^6$Li atoms to the magnetic trap. We implement a two-species magnetic transport system to transport atom clouds from the 3D MOT chamber to the full glass science chamber. The obtained efficiency are 47% for $^{41}$K atoms and 51% for $^6$Li atoms, respectively. To achieve the double degeneracy, we perform a forced RF evaporative cooling for $^{41}$K atoms in the optically-plugged magnetic trap, while $^6$Li atom cloud is cooled sympathetically. The obtained degenerate mixture contains $1.4\times10^5$ $^{41}$K atoms with a condensate fraction of 62% and $5.4\times10^5$ $^6$Li atoms at about 0.25 Fermi temperature. Our work represents an important progress in the study of Bose-Fermi mixtures. Moreover, by replacing the $^{41}$K with $^{40}$K atoms, the fermionic isotope of potassium, the presented apparatus also allows us to study interesting physics of Fermi-Fermi mixtures.

**Acknowledgements**

This work has been supported by the NSFC of China, the CAS, and the National Fundamental Research Program (under Grant No. 2013CB922001). X.-C. Yao acknowledges support from the Alexander von Humboldt Foundation.